# Efficient Toffoli Gates Using Qudits

T.C.Ralph[1], K.J.Resch[2,3] and A.Gilchrist[1]

[1]Centre for Quantum Computer Technology, [2]Department of
Physics, University of Queensland, Brisbane 4072, QLD, Australia,
[3] Institute for Quantum Computing and Department of Physics &
Astronomy, University of Waterloo, Waterloo, ON, N2L 3G1 Canada.
(Dated: November 9, 2018)

The simplest decomposition of a Toffoli gate acting on three qubits requires *five* 2-qubit gates. If we restrict ourselves to controlled-sign (or controlled-NOT) gates this number climbs to six. We show that the number of controlled-sign gates required to implement a Toffoli gate can be reduced to just *three* if one of the three quantum systems has a third state that is accessible during the computation, i.e. is actually a qutrit. Such a requirement is not unreasonable or even atypical since we often artificially enforce a qubit structure on multilevel quantums systems (eg. atoms, photonic polarization and spatial modes). We explore the implementation of these techniques in optical quantum processing and show that linear optical circuits could operate with much higher probabilities of success.

## I. INTRODUCTION

Quantum computing promises major increases in computing power but poses many experimental hurdles to its implementation [1]. Finding more efficient ways to implement quantum gates may allow small scale quantum computing tasks to be demonstrated on a shorter time-scale. A key quantum gate is the Toffoli gate. The Toffoli gate acts on 3 qubits and in conjunction with the Hadamard gate forms perhaps the simplest universal gate set in quantum computing [2].

In this paper we show that the number of two-qubit gates required to implement a Toffoli gate can be reduced by making one of the qubits in the circuit a qutrit (or, in the general case, a qudit). The qutrit nature will only manifest during the gate - after the gate only the qubit levels will be occupied. Remarkably, the additional space afforded by the extra level on one qubit provides a significant reduction in the resource requirements. This reduction is particularly dramatic for optical implementations where systems for applying the envisaged manipulations exist quite naturally. The paper is arranged in the following way. In the next section we introduce the scheme in an abstract, implementation independent, way and demonstrate its increased efficiency. In section III we consider various optical realizations and then summarize and conclude in our final section.

## II. TOFFOLI GATE WITH QUDITS

We begin by showing how a Toffoli gate can be implemented using only three controlled sign (C-S) gates plus single qubit unitaries by allowing one of the qubits to be a qutrit. To our knowledge the most efficient implementation that has been described using only qubits involves five 2-qubit gates [3], [4]. If we restrict ourselves to using only controlled sign (C-S) gates, then six C-S gates plus various single qubit gates are needed [1].

Fig.1 shows the arrangement to implement a Toffoli-Sign (T-S) gate, i.e. a three qubit gate that applies a sign change on one and only one of the state components and the identity is implemented otherwise. It is of no consequence which state component is sign flipped and the flipped component will vary between our various implementations. All such gates are locally equivalent and can be inter-converted with straightforward single-qubit bit flips. The T-S gate becomes a Toffoli gate by placing Hadamards before and after the gate on one of the qubits, which then becomes the target qubit. The two control qubits are labelled as usual with logical states **0** and **1** (where we reserve bold print for indicating logical values). The target is a qutrit for which we label the additional state **2**. We assume C-S gates and Hadamard gates are available. We note that a CNOT gate can be constructed from a C-S gate using a pair of Hadamard gates. We require one additional gate that we label $X_A$, which enables transitions between the **0** and **2** states. The C-S (and CNOT) gate act on the qubit levels in the usual way. If they act on a qutrit in the state **2** then both gates implement the identity regardless of the value of the control qubit. The first operation we perform is to apply an $X_A$ gate to the target qutrit. The $X_A$ gate is defined by the following actions on the basis states: $X_A|\mathbf{0}\rangle \to |\mathbf{2}\rangle$, $X_A|\mathbf{2}\rangle \to |\mathbf{0}\rangle$, $X_A|\mathbf{1}\rangle \to |\mathbf{1}\rangle$. Consider an arbitrary three qubit state

$$\sum_{i,j,k=0,1} \alpha_{i,j,k}|i,j,k\rangle \qquad (1)$$

where $|i,j,k\rangle \equiv |i\rangle_a|j\rangle_b|k\rangle_c$ and the ket subscripts label the different qubits according to Fig.1. After the application of the $X_A$ gate on the qutrit (c) we have

$$\sum_{i,j=0,1} (\alpha_{i,j,0}|i,j,\mathbf{2}\rangle + \alpha_{i,j,1}|i,j,\mathbf{1}\rangle) \qquad (2)$$

We now apply a CNOT gate between b and c to obtain

$$\sum_{i=0,1} (\alpha_{i,0,0}|i,\mathbf{0},\mathbf{2}\rangle + \alpha_{i,0,1}|i,\mathbf{0},\mathbf{1}\rangle +$$

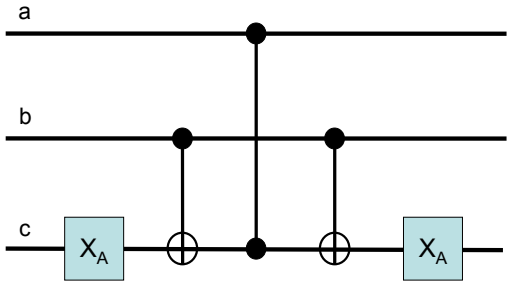

FIG. 1: (Color online) Realization of a T-S gate using two qubits (a and b) and a qutrit (c). CNOT gates (first and last two-qubit gates) operate as normal on the qubit levels and implement the identity if the target is in the qutrit level ($|\mathbf{2}\rangle$). Similarly for the C-S gate (middle two-qubit gate). The $X_A$ gate flips the qutrit between the states $\mathbf{0}$ and $\mathbf{2}$. The sign change occurs on the $|\mathbf{1},\mathbf{0},\mathbf{1}\rangle$ component.

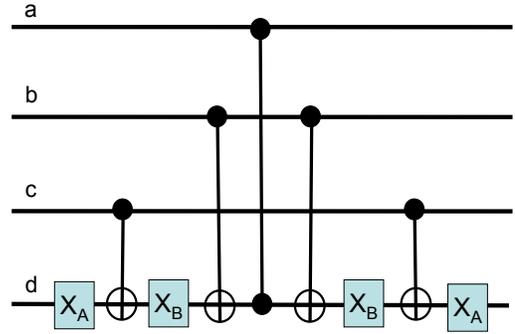

FIG. 2: (Color online) Realization of an n=3 level T-S gate using three qubits (a, b and c) and a ququit (d). CNOT gates operate as normal on the qubit levels and implement the identity if the target is in the ququit levels ($\mathbf{2}$ or $\mathbf{3}$). Similarly for the C-S gate. The $X_A$ gate flips the ququit between the states $\mathbf{0}$ and $\mathbf{2}$. The $X_B$ gate flips the ququit between the states $\mathbf{1}$ and $\mathbf{3}$. The sign change occurs on the $|\mathbf{1},\mathbf{1},\mathbf{1},\mathbf{1}\rangle$ component.

$$\alpha_{i,1,0}|i,\mathbf{1},\mathbf{2}\rangle + \alpha_{i,1,1}|i,\mathbf{1},\mathbf{0}\rangle) \tag{3}$$

Next a C-S gate is applied between $a$ and the $c$ resulting in

$$\begin{aligned}&\alpha_{0,0,0}|\mathbf{0},\mathbf{0},\mathbf{2}\rangle + \alpha_{1,0,0}|\mathbf{1},\mathbf{0},\mathbf{2}\rangle + \alpha_{0,0,1}|\mathbf{0},\mathbf{0},\mathbf{1}\rangle - \\ &\alpha_{1,0,1}|\mathbf{1},\mathbf{0},\mathbf{1}\rangle + \alpha_{0,1,0}|\mathbf{0},\mathbf{1},\mathbf{2}\rangle + \alpha_{1,1,0}|\mathbf{1},\mathbf{1},\mathbf{2}\rangle + \\ &\alpha_{0,1,1}|\mathbf{0},\mathbf{1},\mathbf{0}\rangle + \alpha_{1,1,1}|\mathbf{1},\mathbf{1},\mathbf{0}\rangle\end{aligned} \tag{4}$$

Now a CNOT gate is again applied between $b$ and the $c$ and finally the $X_A$ gate is again applied to the qutrit. The state is then

$$\begin{aligned}&\alpha_{0,0,0}|\mathbf{0},\mathbf{0},\mathbf{0}\rangle + \alpha_{1,0,0}|\mathbf{1},\mathbf{0},\mathbf{0}\rangle + \alpha_{0,0,1}|\mathbf{0},\mathbf{0},\mathbf{1}\rangle - \\ &\alpha_{1,0,1}|\mathbf{1},\mathbf{0},\mathbf{1}\rangle + \alpha_{0,1,0}|\mathbf{0},\mathbf{1},\mathbf{0}\rangle + \alpha_{1,1,0}|\mathbf{1},\mathbf{1},\mathbf{0}\rangle + \\ &\alpha_{0,1,1}|\mathbf{0},\mathbf{1},\mathbf{1}\rangle + \alpha_{1,1,1}|\mathbf{1},\mathbf{1},\mathbf{1}\rangle\end{aligned} \tag{5}$$

as expected for a T-S gate with the sign change implemented on the $|\mathbf{1},\mathbf{0},\mathbf{1}\rangle$ component. This technique can be generalized straightforwardly to higher order n-Toffoli gates (where n is the number of control qubits) by introducing an n+1 level qudit as the target. Fig.2 shows an explicit construction of the next level up, the n=3, T-S gate. In general the number of two-qubit gates required for this method is $2n - 1$. This seems a significant improvement on previous estimates of optimal gate numbers for higher order Toffoli gates. For example reference [5] finds 64 2-qubit gates are required for a 5-Toffoli whilst our qudit construction requires only 9. These results suggest that the computational depth of quantum processing circuits might be significantly reduced by employing these techniques. In the next section we discuss various ways in which the required manipulations can be realized in optics.

## III. OPTICAL IMPLEMENTATIONS

We now turn specifically to an optical encoding and show that for this encoding the manipulations discussed in the previous section have natural physical realizations. First consider dual-rail encoding in which the logical qubit states are given as $|\mathbf{0}\rangle = |1\rangle_h \otimes |0\rangle_v$ and $|\mathbf{1}\rangle = |0\rangle_h \otimes |1\rangle_v$ where $|i\rangle_j$ is the $i$th Fock state of the $j$th optical mode (with plain text reserved for Fock state occupation number). The optical modes are orthogonal and may represent, for example, different polarization or spatial modes. To create a qutrit we simply add an extra mode such that, in the language of the previous section, $|\mathbf{0}\rangle = |100\rangle$, $|\mathbf{1}\rangle = |010\rangle$ and $|\mathbf{2}\rangle = |001\rangle$, where we have simplified our nomenclature such that $|i\rangle_h \otimes |j\rangle_v \otimes |k\rangle_s = |ijk\rangle$, where $s$ labels the new mode. In the following we will show that linear optical elements plus proposed two qubit optical gates are sufficient to implement the operations used in the previous section. We begin by discussing deterministic two qubit gates based on strong non-linearities. We then consider heralded non-deterministic gates based on measurement induced non-linearities. Finally we consider demonstration gates based on post-selected measurement induced non-linearities. In each case we find considerable advantages to our qudit implementation.

### A. Deterministic Gates

It has long been known that a strong cross-Kerr non-linearity enables the implementation of universal optical quantum gates on the encoding we have just introduced

[6]. In particular a $\chi_3$ non-linear medium can be used to induce a cross-Kerr effect between the relevant photon modes. Ideally the cross-Kerr effect will produce the unitary evolution $\hat{U}_K = \exp[i\chi \hat{a}^\dagger \hat{a} \hat{b}^\dagger \hat{b}]$, where $\hat{a}$ is the annihilation operator for one optical mode and $\hat{b}$ another. Consider the schematic set-up of Fig.3. Two polarization encoded qubits are converted into spatial dual rail qubits using polarizing beamsplitters. One mode from each of the qubits is sent through the cross-Kerr material. The operation of this device on an arbitrary two qubit input state is given by the following evolution:

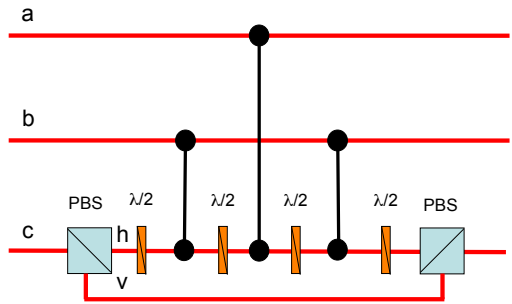

FIG. 4: (Color online) Optical realization of T-S gate. Polarizing beam splitters (PBS) act as $X_A$ gates by accessing an additional spatial mode. The half-wave plates ($\lambda/2$) are set at 22.5 degrees so as to act as Hadamard gates and the C-S gates are implemented as per Fig.3.

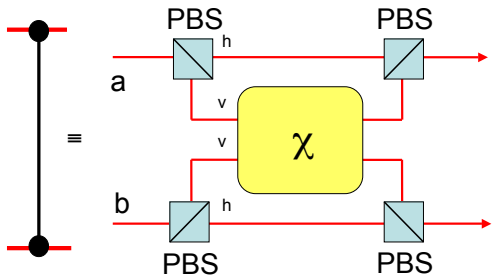

FIG. 3: (Color online) Schematic of the implementation of an optical C-S gate using a strong cross-Kerr non-linearity $\chi$. PBS are polarizing beamsplitters.

$$\begin{aligned}
|\psi\rangle &\to \hat{U}_K |\psi\rangle \\
&= e^{i\chi \hat{a}_v^\dagger \hat{a}_v \hat{b}_v^\dagger \hat{b}_v}(\alpha|10\rangle_a|10\rangle_b + \beta|01\rangle_a|10\rangle_b \\
&\quad + \gamma|10\rangle_a|01\rangle_b + \delta|01\rangle_a|01\rangle_b) \\
&= \alpha|10\rangle_a|10\rangle_b + \beta|01\rangle_a|10\rangle_b \\
&\quad + \gamma|10\rangle_a|01\rangle_b + e^{i\chi}\delta|01\rangle_a|01\rangle_b
\end{aligned} \quad (6)$$

Only when both modes entering the Kerr material are occupied is a phase shift induced. If we now choose the strength of the non-linearity such that $\chi = \pi$, the effect is to flip the sign of one element of the superposition producing a C-S gate. We can introduce a third spatial mode by using polarization rotation and polarizing beamsplitters and hence implement the control gate sequence required in section 1 as shown in Fig.4. Notice, from Eq.6, that this C-S gate implements the identity if both polarisation modes are unoccupied as required. Only 3 $\chi_3$ interactions are required compared with the 5 or 6 that would be needed to implement the Toffoli by the usual qubit gates. Thus the same saving is made as discussed for the abstract case.

Unfortunately Kerr materials with the required strength of non-linearity are not presently available. As a result we now consider non-deterministic implementations based on measurement induced non-linearites.

### B. Heralded Gates

Knill, Laflamme and Milburn have shown that scalable quantum gates can be implemented on optical qubits using linear elements, photon counters and photon sources [7]. In particular, using linear optical elements, photon counters and an entangled photon pair as resources, it is possible to implement the optical C-S gate of the previous section non-deterministically with a probability of success of $1/4$ [8]. Gate success is heralded by the photon counter signature obtained from the entangled pair after interaction with the qubits. If we used the direct method of implementing a Toffoli via a sequence of C-S gates and single qubit unitaries we would require 6 entangled pairs and the probability of success would drop to $(1/4)^6 = 1/4096$. Recently Fiurášek [9] suggested a dedicated scheme requiring only 3 entangled pairs with a probability of success of $1/1065$. His method offered significant reduction in the number of entangled pairs required, but only a modest increase in the success probability. Our method further reduces the number of additional entangled pairs and dramatically improves the success probability. In the following we will adapt our qutrit techniques to linear optics and obtain a heralded Toffili gate requiring only 2 entangled pairs and working with a probability of success of $1/32$.

The proposed set-up is shown in Fig.5. It is similar to the deterministic gate set-up except that the C-S gates are assumed non-deterministic and the final two-qubit gate is replaced with a passive quantum filter. Notice that, because we are also exploiting the polarization degree of freedom on the second spatial mode, the target is now effectively a ququit (a 4-level quantum system). Each 2 qubit gate works with a probability of success $1/4$ and requires an entangled pair for their operation. The filter succeeds with probability $1/2$ and does not require any ancilla qubits for its operation. The gate works in the following way: the state after the second C-S gate (see Fig.5) is

$$\begin{aligned}
&\alpha_{0,0,0}|H,H,vac,H\rangle + \alpha_{1,0,0}|V,H,vac,H\rangle + \\
&\alpha_{0,0,1}|H,H,V,vac\rangle - \alpha_{1,0,1}|V,H,V,vac\rangle + \\
&\alpha_{0,1,0}|H,V,vac,H\rangle + \alpha_{1,1,0}|V,V,vac,H\rangle + \\
&\alpha_{0,1,1}|H,V,H,vac\rangle + \alpha_{1,1,1}|V,V,H,vac\rangle
\end{aligned} \quad (7)$$

conditional on the correct photon counting outcomes from the previous two C-S gates (probability of success $1/16$). Here $H$ and $V$ refer to horizontally and vertically polarized single photon states respectively and $vac$ refers to the vacuum state (both polarization modes unoccupied), i.e $|H\rangle = |10\rangle \equiv |\mathbf{0}\rangle$, $|V\rangle = |01\rangle \equiv |\mathbf{1}\rangle$ and $|vac\rangle = |00\rangle$. The initial state is as defined in Eq.1 and the ordering of the kets corresponds to top to bottom in Fig.5. Instead of using a third C-S gate as in the deterministic scheme, half-wave plates oriented at 22.5 degrees are applied to both target modes leading to the state

$$\begin{aligned}
&\alpha_{0,0,0}|H,H,vac,D\rangle + \alpha_{1,0,0}|V,H,vac,D\rangle + \\
&\alpha_{0,0,1}|H,H,A,vac\rangle - \alpha_{1,0,1}|V,H,A,vac\rangle + \\
&\alpha_{0,1,0}|H,V,vac,D\rangle + \alpha_{1,1,0}|V,V,vac,D\rangle + \\
&\alpha_{0,1,1}|H,V,D,vac\rangle + \alpha_{1,1,1}|V,V,D,vac\rangle
\end{aligned} \quad (8)$$

where $|D\rangle = 1/\sqrt{2}(|H\rangle + |V\rangle)$ and $|A\rangle = 1/\sqrt{2}(|H\rangle - |V\rangle)$. The third and fourth modes are then recombined on a polarizing beam-splitter and the output mode is conditioned on a zero detection at the second output port of the polarizing beamsplitter. The probability of the zero detection is $1/2$ producing the conditional output state

$$\begin{aligned}
&\alpha_{0,0,0}|H,H,H\rangle + \alpha_{1,0,0}|V,H,H\rangle - \\
&\alpha_{0,0,1}|H,H,V\rangle + \alpha_{1,0,1}|V,H,V\rangle + \\
&\alpha_{0,1,0}|H,V,H\rangle + \alpha_{1,1,0}|V,V,H\rangle + \\
&\alpha_{0,1,1}|H,V,V\rangle + \alpha_{1,1,1}|V,V,V\rangle
\end{aligned} \quad (9)$$

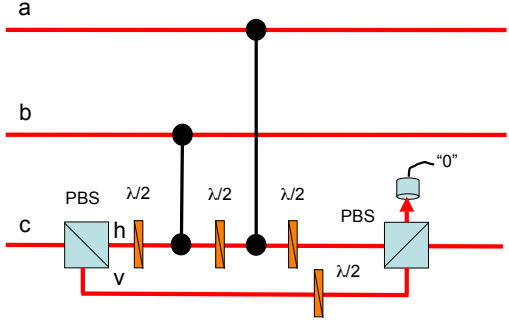

FIG. 5: (Color online) Non-deterministic, but heralded, optical realization of T-S gate. Polarizing beam splitters (PBS) act as $X_A$ gates by accessing an additional spatial mode. The half-wave plates ($\lambda/2$) act as Hadamard gates and the C-S gates are implemented using heralded non-deterministic gates. The sign change occurs on the $|\mathbf{0},\mathbf{0},\mathbf{1}\rangle$ component.

Hence we have produced a T-S gate with the phase flip occurring on the $|H,H,V\rangle$ component. The probability of success is $1/16 \times 1/2 = 1/32$.

The proposed heralded circuit represents a five-photon experiment. Such experiments are feasible, but difficult [10]. In the following section we propose an in principle demonstration requiring only three photons.

### C. Postselected Gates

We now consider the construction of a post-selected gate. By this we mean a gate in which the photons act as their own ancilla, with success heralded by successful detection of a photon for each qubit, so-called coincidence detection. In this way a T-S gate requiring only the 3 photons needed to represent the 3 qubits could be constructed. For example, we could substitute post-selected C-S gates [11, 12] into the circuit of Fig.5. These gates operate with a probability of success of $1/9$ so this would produce a post-selected T-S gate requiring only three photons and working with a probability of success of $1/9 \times 1/9 \times 1/2 = 1/162$. However, it turns out the gate can be optimized for maximum success probability by using the techniques for chaining post-selected C-S gates described in Ref.[13] and in this way achieves a success probability of $1/72$ as we now describe. Fiurášek also considered this problem and proposed a different architecture with a success probability of about $1/133$ [9].

Our proposed set-up is shown in Fig.6. The important part of the circuit is the central string of two interferometers created from the top target mode ($T$). Firstly, if this mode is unoccupied, photons can only emerge in all three output qubits if they remain in their respective modes. In this case no phase flip occurs. If the top target mode is occupied we must consider passage of the photon through the central interferometers. The first interferometer is anti-balanced if the bottom mode of the first control mode, $C1$, is unoccupied. As a result the photon couples completely into the bottom mode of the second interferometer with a phase flip. On the other hand, if the bottom mode of the first control is occupied there will be no phase change at the one third beamsplitter in the first interferometer due to two photon interference (assuming a single photon exits from each port of the beamsplitter). In this case the first interferometer is balanced and the photon will couple completely into the top mode of the second interferometer. Now we turn to the second interferometer. If the top mode of the second control, $C2$, is unoccupied then passage through the bottom mode of the interferometer induces a phase flip. But recall this will only occur if the photon has undergone a phase flip in the first interferometer. Thus if the photon subsequently makes it into the output target mode it will not have undergone a phase flip. Similarly if the target photon couples into the top mode of the second interferometer then it will not have undergone a phase flip in the first interferometer and will not undergo any further phase flip if it subsequently makes it to the output target mode, regardless of the state of $C2$. Finally, if the photon couples into the bottom mode of the second interferometer (meaning that $C1$ had its bottom mode unoccupied, and the photon has picked up a phase flip), and $C2$ has its top mode occupied, two-photon interference at the one third beamsplitter in the second interferometer then leads to no phase flip. Thus if the target photon subsequently makes it to the output, this state component will carry a phase flip. Summarizing, we find that the only state component that will carry a phase flip will be the one in which every qubit was in its logical zero state, where we take occupation of the top mode to be logical zero. The beamsplitter ratios are picked such that all state components have equal probability to lead to an event with a photon appearing in all the output qubits. That probability turns out to be 1/72.

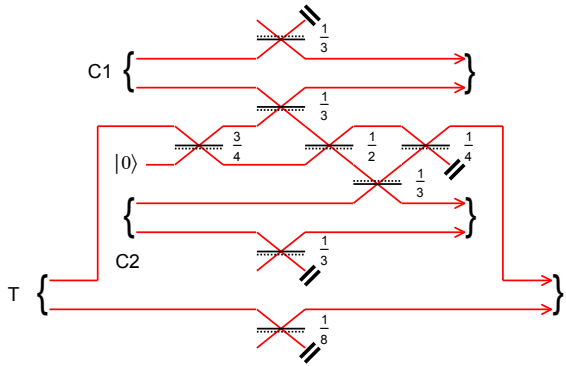

FIG. 6: (Color online) Non-deterministic, post selected, optical realization of a T-S gate. In contrast to the other figures all optical modes are shown explicitly, thus each input qubit is represented by two modes. Note that in the central part of the diagram there are a total of seven modes, due to the introduction of an additional target mode. Beamsplitters are represented as black lines with their reflectivity indicated to the right. The beamsplitters are assumed to be asymmetric, i.e. a phase flip is induced by reflection off one surface but all other components suffer no phase change. The surface for which the phase flip occurs is indicated by a dotted line. If we take occupation of the top mode of each qubit to represent logical zero and occupation of the bottom mode to represent logical one, then the circuit implements a T-S gate in which a phase flip is only applied to the element $|\mathbf{000}\rangle$. The figure is represented, for clarity, as if all modes are spatial. In an experimental realization polarization modes would be utilized where ever possible.

## IV. CONCLUSION

We have discussed the implementation of Toffoli gates in circuits where one of the qubits is allowed to be a qutrit, or more generally a qudit and shown that increased gate efficiency can be achieved. In particular we showed that instead of 6 C-S gates, as required for a qubit only circuit, introducing a qutrit allows this to be reduced to 3 C-S gates. For an $n$-Toffoli the required gate number is $2n-1$ if a $n+1$ level qudit is available. We showed that for deterministic optical quantum processing a natural implementation method could be identified. For non-deterministic optical approaches further simplifications could be identified leading to a heralded Toffoli gate with a probability of success of 1/32 and a post-selected gate with a probability of success of 1/72. These latter results open the door to experimental optical demonstrations of Toffoli gates in the near future and make small scale applications appear more feasible. It is interesting to note that a non-universal set of qubit gates, C-S plus Hadamard (an additional $\pi/8$ gate is required to make this a universal set), becomes universal, Toffoli plus Hadamard, when a single additional quantum level is introduced.

We thank J. L. Dodd, G. J. Pryde and J. L. O'Brien for useful discussions. This work was supported by the Australian Research Council and the DTO-funded U.S.Army Research Office Contract No. W911NF-05-0397.